\documentclass[%
reprint,
superscriptaddress,
 amsmath,amssymb,
 aps,
prb,citeautoscript]{revtex4-1}

\usepackage[T1]{fontenc} 

\usepackage{etoolbox}
\usepackage{graphicx}
\usepackage{dcolumn}
\usepackage{bm}
\usepackage{lipsum}
\usepackage{color}
\usepackage[normalem]{ulem}
\usepackage{simplewick}
\usepackage{qcircuit}
\usepackage{braket}
\usepackage{multirow}
\usepackage{tikz}
\usepackage{xr-hyper}
\usepackage[colorlinks=true,urlcolor=blue,citecolor=blue,linkcolor=blue]{hyperref}
\usepackage{float}
\usepackage[thinlines]{easytable}
\usepackage{booktabs,graphics}

\usepackage{geometry}
\usepackage{pdfpages}

\makeatletter
\AtBeginDocument{\let\LS@rot\@undefined}
\makeatother

\newcommand{\beginsupplement}{%
       \setcounter{table}{0}
       \renewcommand{\thetable}{S\arabic{table}}%
       \setcounter{figure}{0}
       \renewcommand{\thefigure}{S\arabic{figure}}%
    }

\begin{document}

\title{Resource-Efficient Quantum Algorithm for Protein Folding}

\author{Anton Robert}
\affiliation{IBM Research GmbH, Zurich Research Laboratory, S\"aumerstrasse 4, 8803 R\"uschlikon, Switzerland}
\affiliation{PASTEUR, D\'epartement de chimie, \'Ecole Normale Sup\'erieure, PSL University, Sorbonne Universit\'e, CNRS, 75005 Paris, France}

\author{Panagiotis Kl. Barkoutsos}
\affiliation{IBM Research GmbH, Zurich Research Laboratory, S\"aumerstrasse 4, 8803 R\"uschlikon, Switzerland}

\author{Stefan Woerner}
\affiliation{IBM Research GmbH, Zurich Research Laboratory, S\"aumerstrasse 4, 8803 R\"uschlikon, Switzerland}

\author{Ivano Tavernelli}
\email{ita@zurich.ibm.com}
\affiliation{IBM Research GmbH, Zurich Research Laboratory, S\"aumerstrasse 4, 8803 R\"uschlikon, Switzerland}

\date{\today}

\pacs{Valid PACS appear here}

\maketitle

\textbf{
Predicting the three-dimensional (3D) structure of a protein from its primary sequence of amino acids is known as the protein folding (PF) problem.
Due to the central role of proteins' 3D structures in chemistry, biology and medicine applications (e.g., in drug discovery) this subject has been intensively studied for over half a century~\cite{levitt_computer_1975,zwanzig_levinthals_1992,hinds_lattice_1992,shakhnovich_proteins_1995,onuchic_theory_2004,lindorff-larsen_how_2011,hyeon_capturing_2011}.
Although classical algorithms provide practical solutions, sampling the conformation space of small proteins, they cannot tackle the intrinsic NP-hard complexity of the problem~\cite{unger_finding_1993}, even reduced to its simplest Hydrophobic-Polar model \cite{berger_protein_1998}.
While fault-tolerant quantum computers are still beyond reach for state-of-the-art quantum technologies, there
is evidence that quantum algorithms can be successfully used 
on Noisy Intermediate-Scale Quantum (NISQ) computers~\cite{moll_quantum_2017,preskill_quantum_2018}
to accelerate energy optimization in frustrated systems~\cite{mandra_deceptive_2018,king_quantum_2017,denchev_what_2016,streif_comparison_2019}.
In this work, we present a  
model Hamiltonian
with $\mathcal{O}(N^4)$ scaling 
and a corresponding quantum variational algorithm
for the folding of a polymer chain with $N$ monomers on a tetrahedral lattice.
The model reflects many physico-chemical properties of the protein, reducing the gap between coarse-grained representations and mere lattice models.
We use a robust and versatile optimisation scheme, bringing together
variational quantum algorithms specifically adapted to classical cost functions
and evolutionary strategies (genetic algorithms),
to simulate the folding of the 10 amino acid Angiotensin peptide on 22 qubits.
The same method is also successfully applied to the study of the folding of a 7 amino acid neuropeptide using 9 qubits
on an IBM Q 20-qubit quantum computer.
Bringing together recent advances in building gate-based quantum computers with noise-tolerant hybrid quantum-classical algorithms, this work paves the way towards accessible and relevant scientific experiments on real quantum processors.
}

The solution of Levinthal's paradox~\cite{levinthal_are_1968} through a bias search in configuration space~\cite{zwanzig_levinthals_1992} demands a very fine description of the interactions 
in the cellular environment to correctly drive the search in the rugged energy landscape of a protein~\cite{onuchic_toward_1995,onuchic_theory_2004,piana_assessing_2014}. 
GPU-assisted sampling methods of well parametrised coarse-grained models can provide useful insights regarding the protein's native conformation, but folding a small protein in state-of-the-art simulations comes at a very high computational cost~\cite{duan_pathways_1998,lindorff-larsen_how_2011}.
Protein lattice models reduce the conformational space and obviate the high computational cost of an off-lattice exhaustive sampling~\cite{levitt_computer_1975,hinds_lattice_1992}.
Quantum algorithms cannot ignore those simplifications because of curently available quantum resources. 
Perdomo-Ortiz \textit{et al.} paved the way towards the construction of spin Hamiltonian to find the on-lattice heteropolymer's low-energy conformations using quantum devices, but with 
unattainable 
high costs for NISQ computers~\cite{babbush_construction_2012,perdomo_construction_2008}. 
Recently, the amount of resources needed for the simulation of polymer lattice models was reduced, however still maintaining 
an exponential cost in terms of number of qubits and gates needed~\cite{perdomo-ortiz_finding_2012, babej_coarse-grained_2018}.
These methods were used to fold a coarse-grained protein model with $6$ and $8$ amino acid sequences on a 2D and 3D lattice, respectively, using a quantum annealer~\cite{cai_practical_2014}. 
These experiments required $81$ and $200$ qubits and led to a final population of $0.13\%$ and $0.024\%$ for the corresponding ground state structures, using divide and conquer strategies.
More recently, Fingerhuth and coworkers proposed another approach based on the Quantum Approximate Optimization Algorithm (QAOA)~\cite{farhi_quantum_2016} using a problem-specific alternating operator ansatz 
to model protein folding~\cite{fingerhuth_quantum_2018}.
Employing the same model proposed in~\cite{babej_coarse-grained_2018} they succeeded in folding a $4$ amino acid protein model on a 2D square lattice.
\begin{figure}[t!]
\begin{centering}
\includegraphics[width=8.9cm, height=10cm]{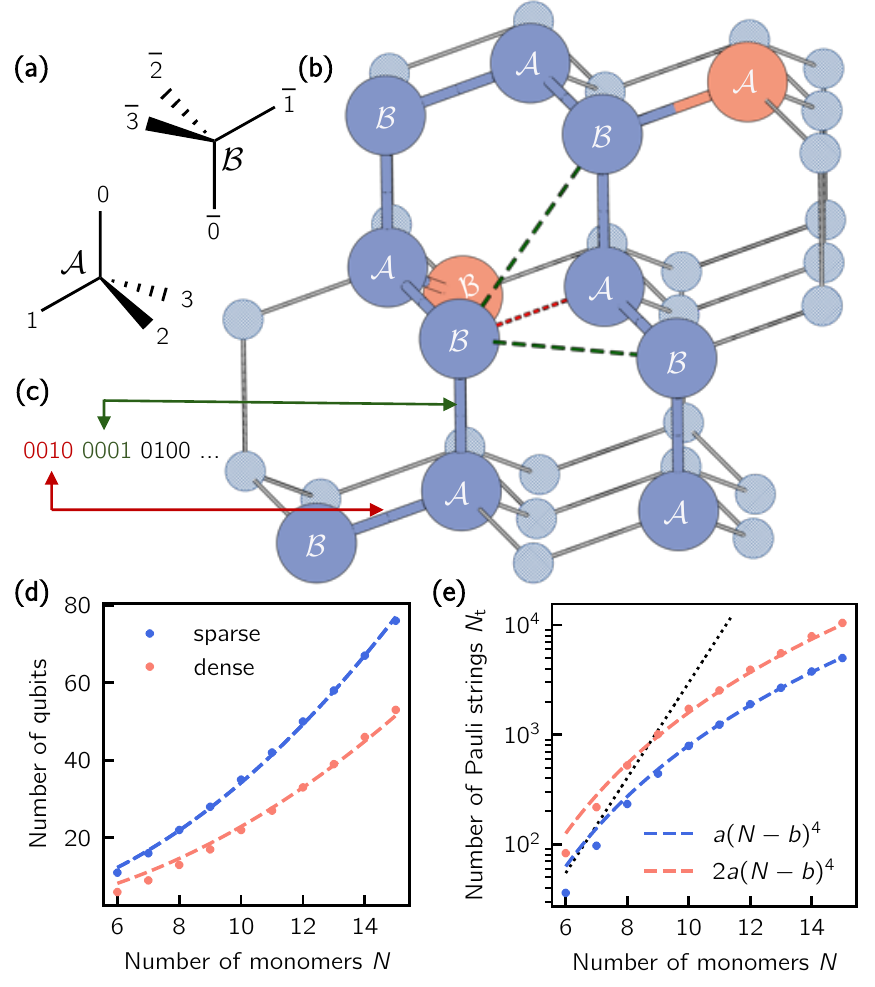}
\par\end{centering}
\caption{
\textbf{Tetrahedral lattice polymer model.} 
\textbf{(a)} Labelling of the coordinate systems at the sub-lattices $\mathcal{A}$ and $ \mathcal{B}$.
\textbf{(b)} Typical polymer conformation (10 monomers). The red and dark green dashed lines represent a subset of inter-bead interactions considered in our model. Side chain beads are shown in orange.
\textbf{(c)} Example of turn encoding. 
\textbf{(d)} Number of qubits required by the sparse (3-local terms) model (blue) and its dense (5-local terms) variant (orange) as a function of the number of monomers.
\textbf{(e)} Number of Pauli strings
with respect to the number of monomers for the sparse and dense encoding models. 
The parameters of the fit are $(a,b)=(0.15,1.49)$. 
The exponential curve (black dotted) is given as a reference.}
\label{fig:Figure1}
\end{figure}

In this work, we present a coarse-grained model for protein folding which is suited to the representation of branched heteropolymers comprised of $N$ monomers on a tetrahedral (or ``diamond'') lattice. 
This choice is motivated by the chemical plausibility of the angles enforced by the lattice ($109.47^o$ for bond angles, $180^o$ or $60^o$ for dihedrals), which allows an all-atom description for a wide range of chemical and biological compounds. 
Given the modest resources of actual quantum devices, a two-centered coarse-grained description of amino acids (backbone and side chain) was used to mimic the protein sequence.
Every monomer is depicted by one or multiple beads that can have a defined number of `color shades' corresponding to different physical properties like hydrophobicity and charge. 
\paragraph*{The configuration qubits.} As for the previous models in  literature, a polymer configuration is grown on the lattice by adding the different beads one after the other and encoding, in the qubit register the different ``turn'' $t_i$ that defines the position of the bead $i+1$ relatively to the previous bead $i$. 
Using a tetrahedral lattice, we distinguish two sets of nonequivalent lattice points $\mathcal{A}$ and $\mathcal{B}$ (see Fig.~\ref{fig:Figure1}). 
At the $\mathcal{A}$ sites, the polymer can only grow along the directions $t_i \in \{0,1,2,3\}$ while at site $\mathcal{B}$ the possible directions are $t_i \in \{\bar{0},\bar{1},\bar{2},\bar{3}\}$. 
Along the sequence, the $\mathcal{A}$ and $\mathcal{B}$ sites are alternated so that we can use the convention that $\mathcal{A}$ (respectively $\mathcal{B}$) sites correspond to even (odd) $i$.
Without loss of generality, the first two turns can be set to $t_1 = \bar{1}$ and $t_2 = 0$ due to symmetry degeneracy. 
To encode the turns, we assign one qubit per axis $t_i=\mathrm{q}_{4i-3}\mathrm{q}_{4i-2}\mathrm{q}_{4i-1}\mathrm{q}_{4i}$ (Fig.~\ref{fig:Figure1}(c)). 
Therefore, the total number of qubits required to encode a conformation $\boldsymbol{\mathrm{q}}_{\text{cf}}$ corresponds to $N_{  \text{cf} }=4(N-3)$. 
If the monomers are described by more than one bead, the same formula holds by replacing $N $ with the total number of beads in the polymer. A denser encoding of the polymer chain using only $2(N-3)$ configuration qubits is presented in the SI.
\paragraph*{The interaction qubits.} 
To describe the interactions, we introduce a new qubit register $\boldsymbol{\mathrm{q}}_{\text{in}}$, composed of $\mathrm{q}^{(l)}_{i,j}$ for each $l^{ \mathrm{th} } $ nearest neighbour ($l$-NN) interaction on the lattice (see red and green dashed lines for $l=1$ and $l=2$ in Fig.~\ref{fig:Figure1}(b)) between beads $i$ and $j$. 
The use of these registers will be explained in connection to the definition of the interaction energy terms. 
The number of qubits constituting the interaction register, $N_{\text{in}}$, is entirely determined by the skeleton of the polymer (i.e. including the side chains), regardless of the beads' color, and scales as $\mathcal{O}(N^2)$.
Note that two $1$-NN beads occupy positions on different sub-lattices ($\mathcal{A}$ or $\mathcal{B}$). 
On the other hand, for $l>1$ all beads of both sub-lattices can potentially interact. 
Given a primary sequence, the pairwise interaction energies $\epsilon^{(l)}_{i,j}$ between the beads at distance $l$ can be arbitrarily defined to reproduce a fold of interest or it can be adapted from pre-existing models, like the one proposed by Miyazawa and Jernigan~(MJ) for $1$-NN interactions~\cite{miyazawa_residueresidue_1996}.
\paragraph*{The Hamiltonian.} The next step
defines the qubit Hamiltonian that describes the energy of a given fold defined by the sequence of beads (fixed) and the encoded turns.
Penalty terms are applied when physical constraints are violated (e.g., when beads occupy the same position on the lattice), and physical interactions (attractive or repulsive in nature) are applied when two beads occupy neighbouring sites or are at distance $l>1$, where $l$ is the number of 
NN in real space. 
The different contributions to the polymer Hamiltonian are therefore (with  $\boldsymbol{\mathrm{q}}=\{ \boldsymbol{\mathrm{q}}_{\text{cf}},\boldsymbol{\mathrm{q}}_{\text{in}}$\}),
\begin{equation}
H(  \boldsymbol{\mathrm{q}} )=
H_{\text{gc}} (\boldsymbol{\mathrm{q}}_{\text{cf}})+ 
H_{\text{ch}}( \boldsymbol{\mathrm{q}}_{\text{cf}} )+
H_{\text{in}}(\boldsymbol{\mathrm{q}})\, .
\label{eq:Hamiltonian}
\end{equation}
The definitions of the geometrical constraint ($H_{\text{gc}}$, which governs the growth of the primary sequence with no bifurcation) and the chirality constraint ($H_{\text{ch}}$, which enforces the correct stereochemistry of the side-chains if present) are given in SI. 
\paragraph*{The interaction energy terms.} 
For each bead $i$ along the sequence the distance to the other beads $j\neq i$ can uniquely be determined by the state
of the $N_{\text{cf}}$ configuration qubits.
To this end, for each pair of beads $(i,j)$ we introduce a four-dimensional vector (Eq.~SI-13), the norm of which uniquely encodes they reciprocal distance $d(i,j)$. 
As an example, we consider the energy contributions for $1$-NN interactions. 
For each pair of beads $(i,j)$ an energy contribution of $\epsilon^{(l)}_{ij}$ is added to $H_{\rm{in}}^{i,j}$ when the distance $d(i,j)=l$. 
However, a contribution of the form $\epsilon^{(l)}_{ij} \, \delta (d(i,j)-l)$ cannot be efficiently implemented as a qubit string Hamiltonian (here $\delta(.)$ stands for the Dirac delta function).
Using the set of contact qubits $\mathrm{q}^{(l)}_{i,j}$ we therefore define an energy term of the form
$\mathrm{q}^{(l)}_{i,j} (\epsilon^{(l)}_{ij} + \lambda (d(i,j)-l))$  
for each value of $l$ and $\lambda \gg \epsilon^{(l)}_{ij}$. 
This definition implies that the contribution $\epsilon^{(l)}_{ij}$ for the formation of the ``interaction'' $(i,j)$ at distance $l$ is only assigned when the contact qubit $\mathrm{q}^{(l)}_{i,j}=1$ and $d(i,j)=l$, simultaneously. For $\mathrm{q}^{(l)}_{i,j}=1$ and $d(i,j)\neq l$ the factor $\lambda$ adds a large positive energy contribution that overcomes the stabilizing energy $\epsilon^{(l)}_{ij}$ (the case of $d(i,j)<l$ is detailed in the SI).
\begin{figure*}[t!]
\begin{centering}
\includegraphics[width=17cm, height=5.5cm]{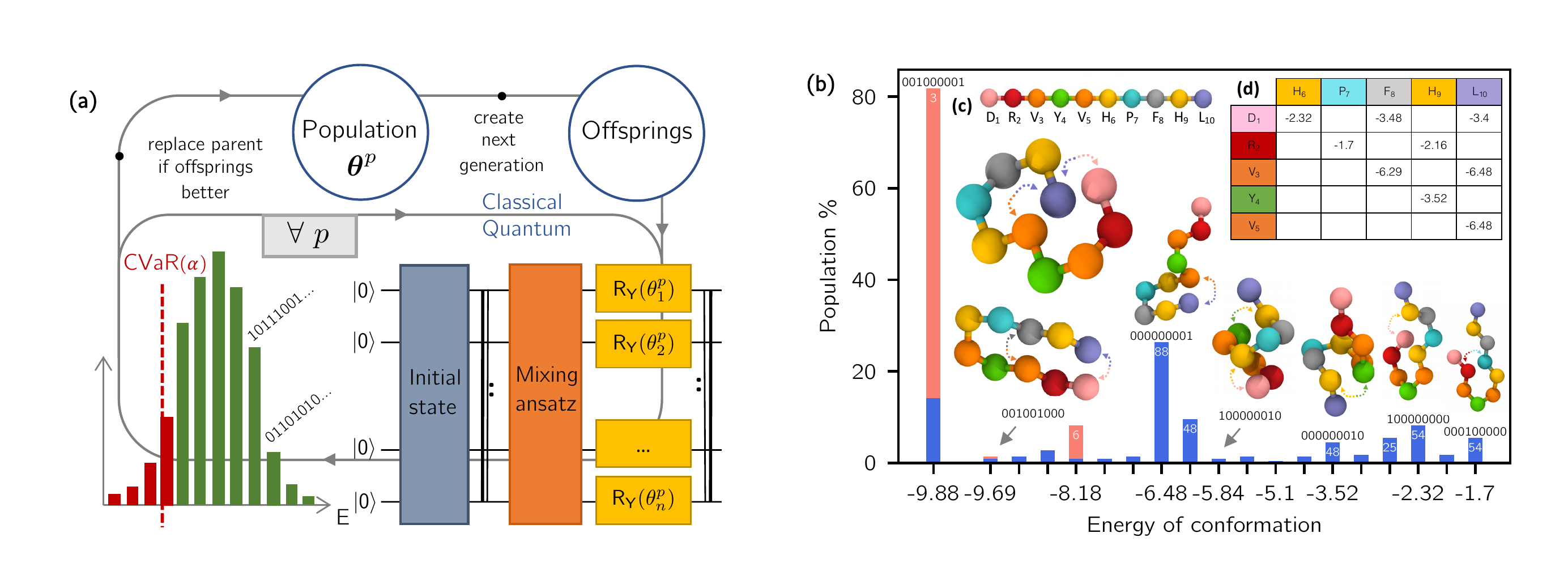}
\par\end{centering}
\caption{
\textbf{(a)} 
\textbf{Schematic representation of the folding algorithm.}
Starting from a random population (up-center) of circuit parameters $\{\boldsymbol{\theta}\}$, every parent, $\boldsymbol{\theta}^p$,  undergoes a parametrized recombination with other individuals according to the procedures detailed in Section Materials and Methods\ref{sec:M&M}. 
The corresponding trial wavefunctions are generated by the quantum circuit as described in the main text and measured to estimate the new CVaR objective function. They constitute the selection criteria of whether to replace a parent by its offspring for the new generation.
\textbf{(b)} 
\textbf{Folding of the ten amino acid Angiotensin peptide.} 
Energy distribution at the convergence of the low-energy folds for the population obtained with the CVaR-VQE algorithm and the DE optimizer. 
The results were obtained using $128$ (blue) and $1024$ measurements (orange).
Simulations were carried out using a realistic parametrization of the noise. 
The binary strings ($ \mathrm{q}_{1,6}\ \mathrm{q}_{1,8}\  \mathrm{q}_{1,10}\ \mathrm{q}_{2,7}\ \mathrm{q}_{2,9}\ \mathrm{q}_{3,8}\ \mathrm{q}_{3,10}\ \mathrm{q}_{4,9}\ \mathrm{q}_{5,10}$) associated with the different bars represent the contact qubits (see main text) that entirely define the conformation energies.
The numbers labelling the bars correspond to the exact degeneracy of the conformations. 
The total percentage of finding low-energy conformations (energy below 0) adds up to $89.5\%$ (small sampling, blue) and $100\%$ (large sampling, orange). 
The fittest individual in the population collapses to the ground state with a probability of $42.2$\% (see SI).
\textbf{(c)} Primary sequence of Angiotensin. 
Each amino acid is assigned a different colour to match its specific physical properties. 
The letters stand for Aspartic-Acid (D), Arginine (R), Valine  (V), Tyrosine (Y), Histidine (H), Proline (P), Phenylalanine (F), and Leucine (L). 
\textbf{(d)} Pairwise interaction matrix for Angiotensin constructed using the MJ model (Table 3 in~\cite{miyazawa_residueresidue_1996}). }
\label{fig:Figure2}
\end{figure*}
Finally, in our model we prevent the simultaneous occupation of a single lattice site by two beads, as discussed in the SI. 
In a nutshell, we only prevent overlaps that occur in the vicinity of an interaction pair. If $\mathrm{q}^{(l)}_{i,j}=1$, we apply penalty functions so that $i$ and $j+1$ cannot overlap when $l=1$, for instance. 
\paragraph*{The folding algorithm.} 
The solution to the folding problem is the ground state of the Hamiltonian $H(\boldsymbol{\mathrm{q}})$ and therefore lies  in the $2^{N_{\text{cf} }}$ dimensional space of the configuration qubits.
To find this solution we prepare a variational circuit, comprising both the configurational and the interaction registers, which is composed by an initialization block with Hadamard gates and parametrized single qubit $R_\text{Y} $ gates followed by an entangling block and another set of single qubit rotations. We denote by $\boldsymbol{\theta}  = ( \boldsymbol{\theta}^{\text{cf} } ,   \boldsymbol{\theta}^{\text{in} }  )$ the set of angles of size $2n$ where  $n=N_{\text{cf}}+N_{\text{ct}}$ is the total number of qubits.
Differently to the quantum mechanical case, for the solution of the `classical problem' (e.g., folding) we do not need an estimate of the Hamiltonian expectation value, but we only require the sampling of the low energy tail of the energy distribution. 
Therefore, the optimization of the angles $\boldsymbol{\theta}$ is performed using a modified version of the Variational Quantum Eigensolver (VQE)~\cite{peruzzo_variational_2014,mcclean_theory_2016} algorithm named Conditional Value-at-Risk (CVaR) VQE or simply CVaR-VQE~\cite{barkoutsos_improving_2019}. 
Briefly, CVaR 
defines an objective function based on the average over the tail of a distribution delimited by a value $\alpha$ (see histogram in Fig.~\ref{fig:Figure2}(a)) which is denoted CVaR$_\alpha (\boldsymbol{\theta}) =  \langle \psi(\boldsymbol{\theta}) \vert H(\boldsymbol{\mathrm{q}})  \vert\psi(\boldsymbol{\theta})\rangle_\alpha $.
Compared to conventional VQE, CVaR-VQE provides a drastic 
speed-up to the optimization of diagonal Hamiltonians as shown in~\cite{barkoutsos_improving_2019}. 
The classical optimization of the gate parameters is performed using a Differential Evolution (DE) optimizer~\cite{storn_differential_1997}, which mimics natural selection in the space of the angles $\boldsymbol{\theta}$.
The optimisation procedure is summarized in Fig.~\ref{fig:Figure2}(a). 
Note that at each step of the optimization, the wavefunctions $|\psi(\boldsymbol{\theta}^p) \rangle$ corresponding to the different individuals  $\boldsymbol{\theta}^p$ (Fig.~\ref{fig:Figure2}(a)) are collapsed during measurement leading to binary strings, which are uniquely mapped to the corresponding configurations and energies. 
We denote by $ \mathbb{P}_{ \text{f} }( p )$ the probability for the individual $p$ to find the $\text{f}^{\text{th}}$ lowest energy fold at convergence. 
\paragraph*{Scaling.}
We define the scaling of the algorithm as the number of terms (or Pauli strings), in the $n$-qubit Hamiltonian $H(\boldsymbol{\mathrm{q}})$ (see also Table I of SI).
\begin{equation}
H(\boldsymbol{\mathrm{q}})=\sum\limits_{\boldsymbol{\gamma}}^ {N_{\mathrm{t}} } h_{\boldsymbol{\gamma}}\bigotimes\limits_{i=1}^{n}\mathrm{q}_{i}^{\gamma_{i}} 
\end{equation}
where  $h_{\boldsymbol{\gamma}}$ real coefficients, $\mathrm{q}_i= (1-\sigma_i^z)/2$ where $\sigma_i^z$ is  $Z$ Pauli matrix with $\gamma_i \in \{0,1\}$, and $N_{\mathrm{t}}$ is the total number of terms.
A thorough investigation of the scaling (see SI)  reveals that the geometrical constraints imposed by the tetrahedral lattice give rise to all possible $2$-local terms within the $N_{\text{cf}}$ conformation qubits. 
Due to the coupling (entanglement) with the interaction qubits the Hamiltonian locality (i.e. the maximum number of Pauli operators different from the identity in 
$H(\boldsymbol{\mathrm{q}})$) is strictly $3$ for the $1$-NN interaction.
Moreover, the scaling is bound by $ N_{\mathrm{t}} \sim N_{\text{in} }   {N_{\text{cf}}\choose{2}} = \mathcal{O} (N^4)$ even for $l$-NN interactions, with $l\geq1$. 
Fig.~\ref{fig:Figure1}(d) and (e) respectively report the scaling of the proposed model and its qubits requirements. 

\begin{figure*}[t!]
\begin{centering}
\includegraphics[width=18.3cm, height=5.5cm]{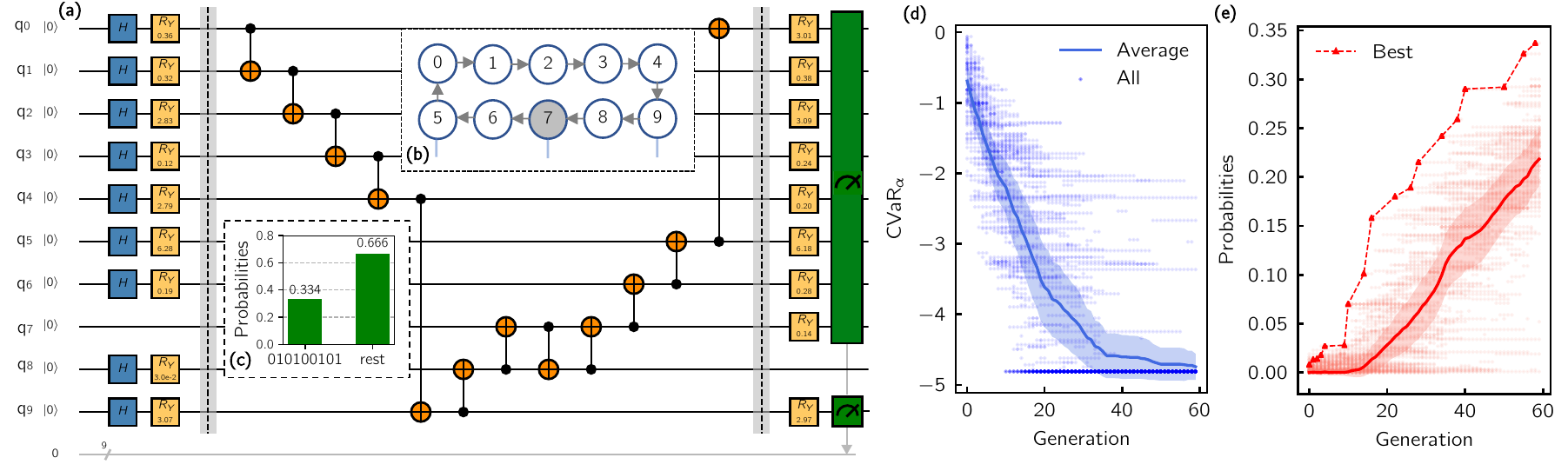}
\par\end{centering}
\caption{
\textbf{Experiment results for the folding of the 7 amino acid neuropeptide.} 
\textbf{(a)} Parametrized quantum circuit for the generation of the configuration space. The optimal set of qubit gate rotations is used to reconstruct the optimal fold. 
\textbf{(b)} Section of the 20 qubits IBM Q Poughkeepsie device used in this experiment. Qubit 7 is used to close the loop by swapping with qubit 8. 
\textbf{(c)} Maximal converged ground state probability for the ground state fold in the population, $\max_p \mathbb{P}_{\text{0}}( p )$. 
\textbf{(d)} Evolution of the ground state energy during the CVaR-VQE minimization (i.e., number of generations) with $\alpha$ parameter set to $5\%$. 
\textbf{(e)} Evolution of the mean probability (over the population ensemble, $\langle \mathbb{P}_{0}(p)\rangle$) and of the best individual probability for the ground state fold, $\max_p \mathbb{P}_{\text{0}}(p)$. 
}
\label{fig:Figure3}
\end{figure*}
\paragraph*{Applications.}
We first apply our quantum algorithm to the simulation of the folding of the 10 amino acid peptide Angiotensin. 
Using our coarse grained model on the tetrahedral lattice the simulation of this system would require 35 qubits, which is computationally intensive.
We therefore introduced a denser encoding of the polymer configuration that requires only 2 qubits per turn $t_i=\mathrm{q}_{2i-1}\mathrm{q}_{2i}$, reducing the total number of qubits to 22. 
This variant generated $5$-local (instead of $3$-local) terms in the qubit Hamiltonian while keeping the total number of Pauli strings within an affordable range for small instances (see Fig.~\ref{fig:Figure1}(d)).
To further reduce the number of qubits we also integrate the side chains with the corresponding bead along the primary sequence and neglect interactions with $l>1$. 
Each bin of the histogram in Fig.~\ref{fig:Figure2}(b) counts, over the population, the occurrence of $ \mathbb{P}_{\text{f}}(p)$ for the minimum energy fold ($\mathrm{f}=0$) and the next 18 folds (histogram bars) given $n_\mathrm{s} = 128$ (blue bars) and  $n_\mathrm{s} = 1024$ (orange bars) measurements of the wavefunctions during the minimization.
More than $80\%$ of the individuals in the final population can generate the minimal conformations after $80$ generations (orange bars), which occurs with a probability $ \max_p \mathbb{P}_{\text{f}}( p ) = 42.2\%$. 
The evolution of the percentage throughout the minimization can be found in the SI. 
By reducing the number of measurements to 128 shots, we obtained a broader spectrum of low energy conformations, which still includes the global minimum but with a lower probability. 
Among the low-energy conformations (with energies below 0), we can clearly identify the formation of an $\alpha$-helix and a $\beta$-sheet (conformations marked with a grey arrow in Fig.~\ref{fig:Figure2}(b)). 
By tuning the interaction matrix (see Fig.S2 in SI), we can foster the formation of secondary structural elements.
The 22-qubit Angiotensin system is still too large for encoding in state-of-the-art quantum hardware.
To this end we used a smaller 7 amino acid neuropeptide with sequence APRLRFY (one-letter coding~\cite{}) that can be mapped to 9 qubits. 
The corresponding CVaR-VQE circuit is shown in Fig.~\ref{fig:Figure3}(a). 
As the entangling block we used a closed-loop of CNOT gates that fits the hardware connectivity of the 20-qubit IBM Q device Poughkeepsie (Fig.~\ref{fig:Figure3}(b)).
The mean CVaR$_\alpha$ energy value of the population as a function of the number of generations shows a robust and smooth convergence towards the optimal fold (Fig.~\ref{fig:Figure3}(d)). 
More importantly, the average probability (Fig.~\ref{fig:Figure3}(e)) of the ground state averaged over the entire population, $\langle \mathbb{P}_{0}(p)  \rangle$, increases monotonically reaching a final value larger than 20\% and with $ \max_p  \mathbb{P}_{0}( p ) $ peaking up at $33\%$ (see Fig.~\ref{fig:Figure3}(c) and (e)).
\paragraph*{Discussion and conclusions.}
In this work, we introduced a quantum algorithm for the solution of the  PF problem on a regular tetrahedral lattice. 
The model Hamiltonain describes a primary coarse-grained protein sequence where each beads represents an amino acid. Side chains can also be modeled by means of an additional bead linked to the main chain. 
The interaction between the amino acids (backbone and side chains) can be extended to $l$-NN (with $l>1$) along the lattice edges. 
This enables the modeling of sophisticated coarse-grained models accounting for Lennard-Jones and Coulombic like interactions. 
We show how the model can correctly reproduce secondary structure elements through the simple adaptation of the imputed contact map.
The number of qubits scales quadratically with the number of amino acids 
, while the number of elements in the Hamltonian scales in $\mathcal{O}(N^4)$.
This implies the use of an unconventional treatment of the overlaps, which are avoided through the addition of penalty terms.
Even though the PF problem is a classical optimization problem, the variational quantum algorithm used, CVaR-VQE, drastically reduces the number of measurements required to minimize the classical cost function (instead of the quantum mechanical average)
and may lead to quantum advantage through the use of entanglement. 
The construction of specific mixing ansatz can drastically speed-up the search in the configuration space even when the ground state is not  entangled but classical~\cite{hadfield_quantum_2019,fingerhuth_quantum_2018}.
The direct connection between qubits and physical properties (configuration and contacts or interactions) allows a rationalisation of initialisation of the qubit states and their entanglement, beyond the simple scheme adopted in this preliminary investigation.

The locality of the Hamiltonian combined with the favourable scaling of the qubit resources and the circuit depth with the number of monomers, make our model the candidate of choice for the solution of the PF problem on NISQ devices and other quantum technologies.
In this work we performed the simulation of the folding of the 10 amino acids protein Angiotensin using a realistic model for the noise of the one and two qubit gate operations.
Furthermore we used a 20 qubit IBM Q processor to compute the folding of a 7 amino acid peptide on 9 qubits, which is 
to our knowledge 
the largest folding calculation on a NISQ device using a variational algorithm. 
The success of this calculation demonstrates the potential of our folding algorithm and opens up new interesting avenues for the use of quantum computers in the optimization of classical cost functions using the CVaR-VQE approach combined with a genetic algorithm for the selection of the best fitting variational parameters.

\section*{\label{sec:contributions}Author's Contributions}
A.R., S.W. and I.T. designed the project. A.R. and P.B. performed the experiments and the simulations. All authors contributed to the analysis of the results and to the writing of the manuscript.

\section*{\label{sec:contributions}Acknowledgements}
I.T. and P.K.B. acknowledge financial support from the Swiss National Science Foundation (SNF) through the grant No. 200021-179312.
A. R. is much obliged to Vladimir Nikolaevitch Smirnov who financially supported his work.
All authors would like to acknowledge support from the IBM Q network and thank the qiskit development team for discussions regarding the development of the software.

IBM, IBM Q, Qiskit are trademarks of International Business Machines Corporation, registered in many jurisdictions worldwide. Other product or service names may be trademarks or service marks of IBM or other companies.

\section*{\label{sec:M&M}Materials and Methods}

The noisy simulations were conducted with $\alpha=1\%$ (resp. $\alpha=0.1\%$) for the $128$ shots (resp. $1024$ shots) simulation with an ``all-to-all`` entangling scheme on Qiskit~\cite{Qiskit}. 
All circuits were constructed with a VQE depth of $m=2$. Given a run with $n$ qubits, the size of the population for the evolutionary algorithm was set to $P=5mn$,  a typical size according to the literature. The selection strategy of the DE algorithm is practically identical to the original ``current-to-best/1/bin'' \cite{das_recent_2016}.


%

\setcounter{equation}{0}
\renewcommand{\theequation}{SI-\arabic{equation}}

\onecolumngrid
\newpage

\section*{Supporting Information for ``Resource-Efficient Quantum Algorithm for Protein Folding''}
\beginsupplement
\section{Lattice model}

We call $i$ the backbone bead of index $i\in\{1,\dots, N\}$ in the primary sequence of the polymer and denote by  $i^{(1)},i^{(2)},...,i^{(s)}$ the beads constituting its side chain. 
Without additional information, discussions about the locality of the Hamiltonian are made with respect to the sparser encoding. 
\subsection{Conformation encoding}

\paragraph*{Sparser Encoding}

The polymer sequence is generated by specifying the series of turns on the lattice, starting from bead $i=1$ (see Fig.~1 of the main text).
Each turn $t_{i}$ is encoded on $4$ qubits $\mathrm{q}_{4i-3}\mathrm{q}_{4i-2}\mathrm{q}_{4i-1}\mathrm{q}_{4i}$, each one representing a direction on the tetrahedral lattice (see Fig.~1 of the main text). 
One and only one qubit of the four will have value one (while the others will be set to zero). 
Accordingly, we encode the turns of the side chain in $i$ as $\mathrm{q}_{i^{(1)}}^{(0)}\mathrm{q}_{i^{(1)}}^{(1)}\mathrm{q}_{i^{(1)}}^{(2)}\mathrm{q}_{i^{(1)}}^{(3)}, \mathrm{q}_{i^{(2)}}^{(0)}\mathrm{q}_{i^{(2)}}^{(1)}\mathrm{q}_{i^{(2)}}^{(2)}\mathrm{q}_{i^{(2)}}^{(3)}...$ 
where the upper index in parenthesis labels the 4 qubits describing the turn of side chain bead $i^{(k)}$ and the
lower indices of the form $k^{(l)}$ label the bead $l$ along the side chain at the main chain position $k$.
Without loss of generality, we choose the first two turns to be $0010$ and $0001$. 
A string of bits defining the entire conformation will have the general form of Eq.~\ref{eq:sparse_encoding} where each 'residue' (main and side chain beads) is embraced within square brackets and the side-chain qubits are within parentheses. Note that the first and the last bead of the main chain (terminal groups) have no side chains. This encoding requires $4(N-1)-8=4(N-3)$ qubits to totally define a conformation. 

\begin{equation}
\left[0010 \right] \left[ 0001(\mathrm{q}_{2^{(1)}}^{(1)}\mathrm{q}_{2^{(1)}}^{(2)}\mathrm{q}_{2^{(1)}}^{(3)}\mathrm{q}_{2^{(1)}}^{(4)}...)\right]\left[ \mathrm{q}_{9}\mathrm{q}_{10}\mathrm{q}_{11}\mathrm{q}_{12}(\mathrm{q}_{3^{(1)}}^{(1)}\mathrm{q}_{3^{(1)}}^{(2)}\mathrm{q}_{3^{(1)}}^{(3)}\mathrm{q}_{3^{(1)}}^{(4)}...)\right]...\left[\mathrm{q}_{4(N-1)-3}\mathrm{q}_{4(N-1)-2}\mathrm{q}_{4(N-1)-1}\mathrm{q}_{4(N-1)}\right]
\label{eq:sparse_encoding}
\end{equation}

\paragraph*{Denser Encoding}
The turn $t_{i}$ is encoded on two qubits $\mathrm{q}_{2i-1}\mathrm{q}_{2i}$.
Accordingly, we will encode the turns in a given side chain by $\mathrm{q}_{i^{(1)}}^{(1)}\mathrm{q}_{i^{(1)}}^{(2)},\mathrm{q}_{i^{(2)}}^{(1)}\mathrm{q}_{i^{(2)}}^{(2)},...$
Without loss of generality we choose the first two turns to be $01$ and $00$. 
If the bead $2$ on the main chain does not bear a side chain, another qubit can be saved without breaking any symmetry ($\mathrm{q}_{6}=1$).
The side chains can be encoded with the same convention as for the sparser encoding. 
A string of bits defining the entire conformation will have the general form of Eq.~\ref{eq:dense_encoding} where the side chain qubits are in parentheses. 
This encoding requires $2(N-1)-4=2(N-3)$ qubits to totally define a conformation. 
\begin{equation}
\left[ 01 \right] \left[ 00(\mathrm{q}_{2^{(1)}}^{(1)}\mathrm{q}_{2^{(1)}}^{(2)}...) \right] \left[ \mathrm{q}_{5}\mathrm{q}_{6}(\mathrm{q}_{3^{(1)}}^{(1)}\mathrm{q}_{3^{(1)}}^{(2)}...) \right] \left[ \mathrm{q}_{7}\mathrm{q}_{8}(\mathrm{q}_{4^{(1)}}^{(1)}\mathrm{q}_{4^{(1)}}^{(2)}...)\right]...\left[ \mathrm{q}_{2(N-1)-1}\mathrm{q}_{2(N-1)}\right] 
\label{eq:dense_encoding}
\end{equation}

\paragraph*{Turn indicator}
It is convenient to introduce an indicator for the axis chosen at turn $t_{i}$. 
We will denote $f_{a}(\mathrm{q}_{2i-1},\mathrm{q}_{2i})=f_{a}(i)$ the function that returns $1$ if the axis $a \in \{0,1,2,3\}$ (see main text, Fig.~1) is chosen at turn $t_{i}$. 
For the denser encoding this function is  given by
\begin{flalign}
 & f_{0}(i)=(1-\mathrm{q}_{2i-1})(1-\mathrm{q}_{2i})\\
 & f_{1}(i)=\mathrm{q}_{2i}(\mathrm{q}_{2i}-\mathrm{q}_{2i-1})\\
 & f_{2}(i)=\mathrm{q}_{2i-1}(\mathrm{q}_{2i-1}-\mathrm{q}_{2i})\\
 & f_{3}(i)=\mathrm{q}_{2i-1}\mathrm{q}_{2i} .\label{eq:indicator} 
\end{flalign}
For the sparser encoding, the expression of $f_{a}(i)$ is trivial:
\begin{flalign}
 & f_{0}(i)=\mathrm{q}_{4i-3}\\
 & f_{1}(i)=\mathrm{q}_{4i-2}\\
 & f_{2}(i)=\mathrm{q}_{4i-1}\\
 & f_{3}(i)=\mathrm{q}_{4i}\label{eq:indicator-1}
\end{flalign}

\paragraph*{Distances}
The shortest spatial distance between two beads is measured as a function of the number of turns separating them along the main chain (see below for the case where the distance is computed between two beads of the side chains).
We define as $n_{a}(i,j)$ (resp. $n_{\bar a}(i,j)$) the number of occurrence of $a \in \{0,1,2,3\}$ (reps. $\bar{a} \in \{\bar{0}, \bar{1}, \bar{2}, \bar{3}\}$) along the sequence separating $i$ and $j$ with $j>i$. All intermediate beads will be labeled by $k=i,\dots,j-1$.
We assume that the polymer starts with a bead on the sub-lattice $\mathcal{B}$. Due to the alternation of the two sub-lattices, all sites of the sub-lattice $\mathcal{A}$ ($\mathcal{B}$) will be labelled with even (odd) integers. 
Let's denote $\Delta n_{a}(i,j)=n_{a}(i,j)-n_{\bar a}(i,j)$ with $n_{a}(i,j)$ (resp. $n_{\bar{a}}(i,j)$) the number of occurrence of $a$ (reps. $\bar{a}$) in the sequence between bead $i$ and $j$. 
We then have
\begin{equation}
\Delta n_{a}(i,j)=\sum_{k=i}^{j-1}(-1)^{k}f_{a}(k)\label{eq:distance_beads}
\end{equation}
The distance between beads that are placed on side chains, say $i^{(s)}$ ($s^{th}$ bead of side chain at position $i$ of the main chain) and $j^{(p)}$ ($p^{th}$ bead of side chain at position $j$) with $j>i$ is obtained by first considering the distance between $i$ and $j$ and then add or remove the contributions of side chain qubits. 
Because $j>i$, the side chain qubits of $j$ (resp. $i$) are taken into account with a plus (resp. minus) sign. 
Moreover, the direction is defined by the parity of $i$ and $j$.
The generalized expression for $\Delta n_{a}$ is therefore
\begin{equation}
\Delta n_{a}(i^{(s)},j^{(p)})=\Delta n_{a}(i,j)+\sum_{l=1}^{p}(-1)^{l}f_{a}(j^{(l)})-\sum_{m=1}^{s}(-1)^{m}f_{a}(i^{(m)})\label{eq:distance_side_chain_beads}
\end{equation}
For the calculation of the actual distance, is is convenient to introduce the four dimensional vector $\mathbf{x}(i,j)$ defined as
\begin{equation}
\mathbf{x}(i,j)=\begin{pmatrix}\Delta n_{1}(i,j)\\
\Delta n_{2}(i,j)\\
\Delta n_{3}(i,j)\\
\Delta n_{4}(i,j)
\end{pmatrix}\label{eq:vector}
\end{equation}
such that $d(i,j) = ||\mathbf{x}(i,j)||^{2}_{2}= \sum_a \Delta n_{a}(i,j)^2 $. It is crucial to note that on the tetrahedral lattice there is a bijective map between the values of $d(i,j)$ and the Euclidean (through-space) distances $r_{ij}$ in lattice bond units, 
\begin{equation}
\begin{array}{cccc}
d(i,j)=0\Rightarrow r_{ij}=0\\
d(i,j)=1\Rightarrow r_{ij}=1\\
d(i,j)=2\Rightarrow r_{ij}=2\sqrt{\frac{2}{3}}\simeq1.63\\
d(i,j)=3\Rightarrow r_{ij}=\sqrt{\frac{11}{3}}\simeq1.91\\
d(i,j)=4\Rightarrow r_{ij}=\frac{4}{\sqrt{3}}\simeq2.31\\
d(i,j)=5\Rightarrow r_{ij}=\sqrt{\frac{19}{3}}\simeq2.52
\\...&
\end{array}\label{eq:nearest neighbours}
\end{equation}

\subsection{Construction of local Hamiltonians}

\paragraph*{Sparser encoding Hamiltonian} 
For the sparser encoding, we need to impose that one and only one of the four qubits that define a turn is equal to one.
On a quantum circuit, this can be achieved easily by using a valid initialization of the qubits together with gate operations that conserved the number of 1's in each set of 4 qubits that encodes a turn. 
When this is not possible, a penalty function as show in Eq~\ref{eq:optional} with a large positive $\lambda$ can be used to impose this constraint.
\begin{equation}
H_{\text{optional}}=\sum_{i=3}^{N-1}\lambda(\mathrm{q}_{4i-3}+\mathrm{q}_{4i-2}+\mathrm{q}_{4i-1}+\mathrm{q}_{4i}-1)^{2}
\label{eq:optional}
\end{equation}

\paragraph*{Growth constraint.}
Several constraints are needed to prevent the growth of the chain towards unphysical geometries  (e.g., to prevent that the chain (main or side) at side $i$ folds back into itself). To this end, we compute function $T(i,j)$ defined as
\begin{equation}
T(i,j)=\sum_{a=\{0,1,2,3\}}f_{a}(i)f_{a}(j)\label{eq:same_turn}
\end{equation}
for each pair of beads $i$ and $j$. $T(i,j)$ returns a 1 if and only if the turns $t_i$ and $t_j$ are along the same axis ($a$ and $\bar{a}$, respectively).
Note that $T(i,j)$ is composed of $2$-local terms for the sparse encoding.
Firstly, we need to eliminate sequences where the same axis ($a$ and $\bar a$) is chosen twice in a row (e.g. $2\bar{2}$) since this will give rise to a chain folding back into itself. 
To this end we apply the following penalty term 
\begin{equation}
H_{\mathrm{gc}}=\sum_{i=3}^{N-1}\lambda_{\text{back}}T(i,i+1)
\end{equation}
with large positive $\lambda_{\text{back}}$.
Note that we can easily control the number of 2-local terms appearing in the sum (linear number of two local terms). 
In the general case, for a degree of branching $s>1$, similar  terms need to be added to prevent the overlap of two consecutive bonds within the side chains.

\paragraph*{Chirality constraints}
Natural polymers have a well defined chirality that has to be imposed in our model. 
In proteins, the position of the side chains at the insertion point with the main chain determines the chirality of each  residue. 
The position of the first side chain bead $i^{(1)}$ on the main bead $i$ is imposed by the choice of the (main chain) turns $t_{i-1}$ and $t_{i}$.

To enforce the correct chirality, we add a constraint $H_{\mathrm{ch}}$ to the Hamiltonian.
The required (expected) chirality at $i^{(1)}$ is encoded in the function $f_{a}^{\text{ex}}(i^{(1)})$ which is a  function of $f_{a}(i-1)$ and $f_{a}(i)$ defined in Eq.\ref{eq:indicator}.

\begin{table}
\begin{centering}
\renewcommand{\arraystretch}{1.2}
\begin{tabular}{c|c|c|c|c}
\toprule
$\begin{array}{cc}
 & t_{i}\\
t_{i-1}
\end{array}$ & 0 & $1$ & $2$ & $3$\tabularnewline
\hline 
$\bar{0}$ &  & $2$ & $3$ & $1$\tabularnewline
\hline 
$\bar{1}$ & $3$ &  & $0$ & $2$\tabularnewline
\hline 
$\bar{2}$ & $1$ & $3$ &  & $0$\tabularnewline
\hline 
$\bar{3}$ & $2$ & $0$ & $1$ & \tabularnewline
\bottomrule
\end{tabular}\hspace{2cm}%
\begin{tabular}{c|c|c|c|c}
\toprule
$\begin{array}{cc}
 & t_{i}\\
t_{i-1}
\end{array}$ & $\bar{0}$ & $\bar{1}$ & $\bar{2}$ & $\bar{3}$\tabularnewline
\hline 
$0$ &  & $\bar{3}$ & $\bar{1}$ & $\bar{2}$\tabularnewline
\hline 
$1$ & $\bar{2}$ &  & $\bar{3}$ & $\bar{0}$\tabularnewline
\hline 
$2$ & $\bar{3}$ & $\bar{0}$ &  & $\bar{1}$\tabularnewline
\hline 
$3$ & $\bar{1}$ & $\bar{2}$ & $\bar{0}$ & \tabularnewline
\bottomrule
\end{tabular}
\par\end{centering}
\caption{Value of the turn parameter $t_{i^{(1)}}$ imposed by chirality constraint as a function of the values $t_{i-1}$
and $t_{i}$ for the case $i$ even (i.e., belonging the  the sub-lattice $\mathcal{A}$). The case $i$ odd is identified
by transposing the matrix. }
\label{tab:chirality}
\end{table}

Analyzing all possible cases, we can construct the truth-table given in Table~\ref{tab:chirality}, which uniquely define $f_{a}^{\text{ex}}(i^{(1)})$
for $a\in\{0,1,2,3\}$.
To this end, we first need to define an indicator for the parity, 
$g_{i}$, which is 0 if $i$ is even and 1 otherwise. 
We then write $g_{i}=\frac{1-(-1)^{i}}{2}$. 
Using the information  in Table~\ref{tab:chirality} and the definition of $g_{i}$, we 
obtain 
\begin{align}
f_{0}^{\text{ex}}(i^{(1)}) & =(1-g_{i})\Bigl(f_{1}(i-1)f_{2}(i)+f_{2}(i-1)f_{3}(i)+f_{3}(i-1)f_{1}(i)\Bigr)\\
 & +g_{i}\Bigl(f_{1}(i)f_{2}(i-1)+f_{2}(i)f_{3}(i-1)+f_{3}(i)f_{1}(i-1)\Bigr)\\
f_{1}^{\text{ex}}(i^{(1)}) & =(1-g_{i})\Bigl(f_{0}(i-1)f_{3}(i)+f_{2}(i-1)f_{0}(i)+f_{3}(i-1)f_{2}(i)\Bigr)\\
 & +g_{i}\Bigl(f_{0}(i)f_{3}(i-1)+f_{2}(i)f_{0}(i-1)+f_{3}(i)f_{2}(i-1)\Bigr)\\
f_{2}^{\text{ex}}(i^{(1)}) & =(1-g_{i})\Bigl(f_{0}(i-1)f_{1}(i)+f_{1}(i-1)f_{3}(i)+f_{3}(i-1)f_{0}(i)\Bigr)\label{eq:expected_chirality}\\
 & +g_{i}\Bigl(f_{0}(i)f_{1}(i-1)+f_{1}(i)f_{3}(i-1)+f_{3}(i)f_{0}(i-1)\Bigr)\\
f_{3}^{\text{ex}}(i^{(1)}) & =(1-g_{i})\Bigl(f_{0}(i-1)f_{2}(i)+f_{1}(i-1)f_{0}(i)+f_{2}(i-1)f_{1}(i)\Bigr)\\
 & +g_{i}\Bigl(f_{0}(i)f_{2}(i-1)+f_{1}(i)f_{0}(i-1)+f_{2}(i)f_{1}(i-1)\Bigr)
\end{align}
The penalty term to impose the right chirality, $H_{\mathrm{ch}}$, will then the penalize the structures for which $f_{a}(i^{(1)})$ differs from $f_{a}^{\text{ex}}(i^{(1)})$,
\begin{equation}
H_{\mathrm{ch}}=\lambda_{\text{chirality}}\sum_{a}\sum_{i=2}^{N-1}\biggl(1-f_{a}(i^{(1)})\biggr)f_{a}^{\text{ex}}(i^{(1)})\label{eq:chirality} \, ,
\end{equation}
for large and positive values of $\lambda_{\text{chirality}}$.
Note that this term actually only contains linear number of 3-local terms for the denser encoding (5-local for the sparser encoding).

\subsection{Construction of $H_{\text{in}}$ }

\paragraph*{Definitions.} 
We can decompose the interaction Hamiltonian into different Hamiltonian terms $H^{(l)}$ corresponding to $l^th$ nearest neighbour interactions,
\begin{equation}
H_{\text{in}}=H_{\text{in}}^{(1)}+H_{\text{in}}^{(2)}+H_{\text{in}}^{(3)} + \dots
\end{equation}
To keep the number of terms in $H_{\text{in}}$ as low as possible we will need to exploit  properties of the tetrahedral lattice. 
In this section, we only consider interactions between beads $i$ and $j$ with $j > i+2$.
Indeed, interactions between beads that are nearest or second nearest neighbour in the primary sequence are present in every conformations so considering them will not help discriminating the folds. Note also that beads that are separated by less than 5 bonds cannot be nearest neighbour (1-NN) on the lattice.
These beads can indifferently represent main chain beads as well as side chain beads since only the absolute distance measure in bond units along the polymer chain matters (e.g. $j^{(1)}-i\equiv j-i+1$ or $j^{(1)}-i^{(1)}\equiv j-i+2$). 
First, we observe that the parity of $j-i$ corresponds to the parity of $d(i,j)$. 
Then, since we alternatively choose among positive and negative directions to encode a conformation, we have
\begin{equation}
(j-i)=\begin{cases}
0\mod2 & {\rm  if} \ \sum_{a}\Delta n_{a}(i,j)=0\\
1\mod2 & {\rm if  } \ \sum_{a}\Delta n_{a}(i,j)=(-1)^{i}
\end{cases}\label{eq:sum_delta} \, .
\end{equation}
In addition, we introduce a notation for the set of first nearest neighbor beads to a given bead $i$, which will be denoted $\mathcal{N}(i)$. 
For example, $\mathcal{N}(i)=\{i-1,i+1,i^{(1)}\}$ if $i$ is not at the end of the polymer chain or $\mathcal{N}(i^{(1)})=\{i\}$ when limited to one side chain bead per monomer. 
Finally, we define a new set of two-index quantities  $\mathrm{q}^{(l)}_{ij},\mathrm{q}^{(l)}_{i^{(1)}j},\mathrm{q}^{(l)}_{i,j^{(1)},}\mathrm{q}^{(l)}_{i^{(1)}j^{(1)}}$ that encode the presence of an interaction of order $l$ between beads $i$ and $j$ (i.e, $i$ and $j$ are l-NN). 

\paragraph*{First nearest neighbour interactions: $H^{(1)}$. }
The quantity $\mathrm{q}^{(1)}_{ij}$ describes a contact between a pair of beads $i$ and $j$ that belong to different lattices and that can be any far apart along the polymer chain. 
If this contact occurs, the energy will be modified by the contact energy $\epsilon^{(1)}_{ij}$. The corresponding term in the Hamiltonian is then $\mathrm{q}^{(1)}_{ij}\epsilon^{(1)}_{ij}$. 
However, since we keep the configuration and the interaction qubit registers independent from each other, before assigning the energy contribution  $\epsilon^{(1)}_{ij}$ we need to make sure that the two beads are indeed at a distance of $1$.
The correct energy term is therefore 
$\mathrm{q}^{(l)}_{ij}=1$ and reads $\mathrm{q}^{(1)}_{ij}\lambda_{1}(d(i,j)-1)$, which implies that we apply either the stabilization energy $\epsilon^{(1)}_{ij}$ when both $\mathrm{q}^{(1)}_{ij}=1$ and $d(i,j)=1$, or a penalty contribution weighted by the positive factor $\lambda_{1}$. 
Here $d(i,j)$ cannot be equal to 0 because the two beads must be on different lattices.
Note that if $r \in \mathcal{N}(j)$ and $i$ and $j$ are first nearest neighbours, $r$ can either overlap with $i$ or be at a distance of $2$ from $i$. 
In order to avoid the overlap of different beads, we penalize  the occurrence of the first option by applying a penalty function.
This constraint can be written as $\mathrm{q}^{(1)}_{ij}\lambda_{2}(2-d(i,j))$ with $\lambda_{2}$ being a large positive value; it  will prevent the formation of a local overlaps between polymer beads in the vicinity of a nearest neighbor contact. 
Summing over all the possible contacts between backbone beads, we obtain the general form of  the first nearest neighbour interaction Hamiltonian
\begin{equation}
H^{(1)}=\sum_{i=1}^{N-4}\sum_{\substack{j\geq i+5\\ j-i=1\mod2 }} ^{N}  h^{(1)}_{i,j} 
\end{equation}
 with 
\begin{equation}
h^{(1)}_{i,j} = \mathrm{q}^{(1)}_{ij} \biggl(\epsilon^{(1)}_{ij}+\lambda_{1}(d(i,j)-1)+\sum_{r\in\mathcal{N}(j)}\lambda_{2}(2-d(i,r))+\sum_{m\in\mathcal{N}(i)}\lambda_{2}(2-d(m,j))\biggr)
\label{eq:H_BBBB}
\end{equation}
The choice of the ratio $\lambda_{1}/\lambda_{2}$ has to be made wisely. 
In fact, we need that the term within parentheses in Eq.~\eqref{eq:H_BBBB} is always positive when $i$ and $j$ are not in contact. 
Since $\vert \mathcal{N}(i) \vert \le 3$ and $ -(j-i+1) < d(i,r) - d(i,j) < (j-i+1) $ when $r \in \mathcal{N}(j)$, we need $\lambda_{1}>6(j-i+1)\lambda_{2} + \epsilon^{(1)}_{ij}$.
The penalty values $\lambda_{1}$ and $\lambda_{2}$ 
therefore depend on $i$ and $j$.

\paragraph*{Second nearest neighbour interactions: $H^{(2)}$. }

\begin{figure*}
\begin{centering}
\includegraphics[width=17cm, height=4cm]{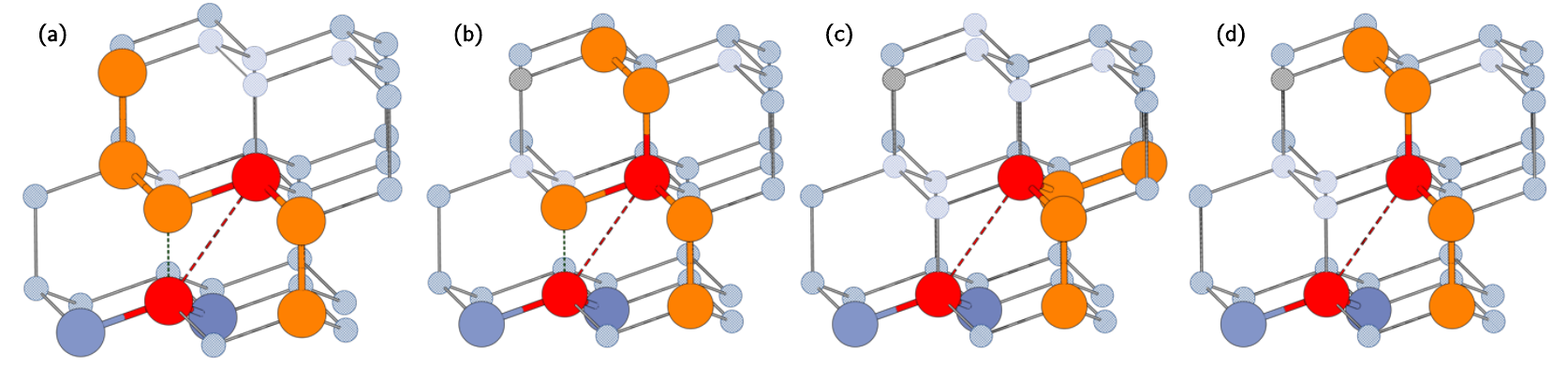}
\par\end{centering}
\caption{Different contributions to $H^{(2)}$: (a) and (b) two different configurations contributing to $H^{(2)}[1]$, (c) a configuration contributing to $H^{(2)}[3,3]$ and (d) a configuration contributing to $H^{(2)}[3,5]$ (or $H^{(2)}[5,3]$ depending on ordering of the beads $r$ and $m$). }
 \label{fig:interactions}
\end{figure*}

In order to take into account second nearest neighbours interactions between beads $i$ and $j$ (which by definition occupy the same sub-lattice), we need to introduce a new set of interaction qubits:  $\mathrm{q}^{(2)}_{ij}$. 
In this case, there are multiple possible configurations that lead to 2-NN interactions, so multiple interaction qubits are needed for each pair $(i,j)$. 
We deal with these different cases by splitting the Hamiltonian as follows
\begin{equation}
H^{(2)}=H^{(2)}[1]+H^{(2)}[3,3]+H^{(2)}[3,5]+H^{(2)}[5,3] \, .
\end{equation}
$ H^{(2)}[1] $ deals with the case in which a given configuration leads to a direct contact between beads $r$ and $i$ or $m$ and $i$ where $(r,m) \in \mathcal{N}(j)$ (see Fig~\ref{fig:interactions}),
\begin{equation}
H^{(2)}[1]=\sum_{i=1}^{N-4}\sum_{\substack{j\geq i+4\\
j-i=0\mod2
}
}^{N} h_{ij}^{(2)}[1]
\end{equation}
where 
\begin{equation}
h_{ij}^{(2)}=\sum_{r\in\mathcal{N}(j)}\text{q}_{i,r}^{(1)}\epsilon_{i,j}^{(2)}
\end{equation}

\pagebreak 
The remaining configurations can be classified into the following classes depending of the distances between bead $i$ and the beads $r$ and $m$, with $r,m \in \mathcal{N}(j)$ (see Fig~\ref{fig:interactions}): 
\begin{itemize}
\item $  [3,3] : d(i,r)=3 $ and  $d(i,m)=3 $ 
\item $  [3,5] : d(i,r)=3 $ and  $d(i,m)=5 $
\item $  [5,3] : d(i,r)=5 $ and $d(i,m)=3  $. 
\end{itemize}
We thus introduce three second contact qubits $\mathrm{q}_{ij}^{(2)}[3,3],\mathrm{q}_{ij}^{(2)}[3,5],\mathrm{q}_{ij}^{(2)}[5,3]$ and use the same method to build the Hamiltonian. For instance, 
\begin{equation}
H^{(2)} [3,5] = \sum_{i=1}^{N-4}\sum_{\substack{j\geq i+4\\
j-i=0\mod2
}
}^{N}  h_{ij}^{(2)}[3,5]
\label{eq:H_35}
\end{equation}
with
\begin{equation}
h_{ij}^{(2)}[3,5] = (1-\mathrm{q}^{(1)}_{i,r})(1-\mathrm{q}^{(1)}_{i,m})\mathrm{q}_{ij}^{(2)}[3,5]\biggl(\epsilon_{ij}^{(2)}+\lambda_{2}(d(i,j)-2)+\lambda_{3}(3-d(i,r))+\lambda_{5}(5-d(i,m))\biggr)
\end{equation}
Note that we also add a check for $\mathrm{q}^{(1)}_{i,r} \neq 1 $ and $\mathrm{q}^{(1)}_{i,m} \neq 1$ to avoid double counting of the interaction energy $\epsilon_{i,j}^{(2)}$.
Also in this case, we choose $\lambda_{2}, \lambda_{3} ,\lambda_{5}$  so that the term in the parenthesis is always positive if $ d(i,j)\neq 2$. 
Additional qubits are necessary when dealing with side chains. The nature of these additional terms can be easily elaborated once the gist of the method is understood. 
To take into account $l$-NN interactions with $l > 2$ we follow the same method of enumerating all the possible configurations giving rise to the specific interactions.
The number of required qubits will evolve exponentially with $l$ but still quadratically with $N$ if truncated to a cutoff maximum value.  The interaction energy can also depend on the configuration of interaction.

\newcommand{\ra}[1]{\renewcommand{\arraystretch}{#1}}
\setlength\extrarowheight{0.5pt}
\begin{table*}[h!]
\ra{1.3}
\begin{center}
\begin{tabular}{c|c|c|c|c|c}
\toprule
 & Perdomo \textit{et al.} \citep{perdomo_construction_2008_si} & Perdomo \textit{et al.} \citep{perdomo-ortiz_finding_2012_si} & Babbush \textit{et al.} \citep{babbush_construction_2012_si}  & Babej \textit{et al.} \citep{babej_coarse-grained_2018_si, fingerhuth_quantum_2018_si}  & This model \tabularnewline
\hline
Model & Hydrophobic-Polar (HP) & Coarse-Grained & Coarse-Grained &  Coarse-Grained &  Coarse-Grained \tabularnewline
\hline
Lattice & all & all & all & all & Tetrahedral \tabularnewline
\hline
Types & $2$ & $\infty$ & $\infty$ & $\infty$ & $\infty$ \tabularnewline
\hline
Interactions & Nearest & Nearest & Nearest & Nearest & $l^{\mathrm{th}}$ Nearest \tabularnewline
\hline
Locality & $\log_{2}N$ & $N$ &  $4$ &  $ N $ & $l+2$ \tabularnewline
\hline
Qubits & $N\log(N)$ & $N^2\log(N)$  & $N^{3}\log(N)$ & $N$ & $N^{2} \exp(l)$ \tabularnewline
\hline
Scaling & $N^{8}$ &  $\exp(N)$ &  $ N^{12} \log^4(N) $  & $\exp(N)$ &  $N^{4}$ \tabularnewline
\hline
Experiment & No &  D-Wave &  No  & Rigetti QPU & IBM QPU \tabularnewline
\bottomrule
\end{tabular}
\caption{\textbf{Comparison of existing models for 3D protein native structure
prediction using quantum algorithms.} 
The sore point of the existing models  \citep{perdomo_construction_2008_si,babbush_construction_2012_si,perdomo-ortiz_finding_2012_si,babej_coarse-grained_2018_si, fingerhuth_quantum_2018_si} is the locality of the generated Hamiltonian and/or the required number of ancillas (qubit resources).
In `digital' quantum computers (last two columns), the coupling between more than two qubits  can  be decomposed into two-qubit controlled operations at the cost of an increased circuit depth.
On the other hand, the architecture of analog quantum annealers prevents the optimization of non-Ising like Hamiltonian (i.e., with $k$-body interactions between qubits). In fact, locality-reduction to an isospectral Ising Hamiltonian ($k=2$) requires the introduction of additional ancillas qubits and the use of complex mapping schemes~\citep{Kempe2006_si,Barkoutsos_fermionic_2017_si} (first three columns).
Since $k$-local terms require $k-2$ ancillas to be embedded, the resource requirements for both quantum approaches (digital and analog) scales exponentially~\citep{babbush_construction_2012_si} with the polymer size if the locality of the model depends on the polymer length, $N$, for coarse-grained models. 
While for the model presented in this work the locality is independent from $N$ this is not the case for most of the proposed Hamiltonians in the literature (Perdomo~\textit{et al.}~\citep{perdomo_construction_2008_si}, Perdomo~\textit{et al.}~\citep{perdomo-ortiz_finding_2012_si}, and Babej~\textit{et al.}~\citep{babej_coarse-grained_2018_si, fingerhuth_quantum_2018_si}).
The corresponding scalings and the localities are defined in the text. For our model, they are obtained after constructing the Hamiltonian on Qiskit~\citep{Qiskit_si}.
The values recorded are obtained without use of any locality reduction schemes.
}
\label{table:comparison}
\end{center}
\end{table*}

\newpage 

\section{Contact Map}

Contact map for the stabilization of two different secondary structure elements: $\beta$-sheet (left) versus $\alpha$-helix (right). 
Simulations performed with these contact maps reproduce the correct minimal energy structures (see corresponding cartoons).
The $\alpha$-helix is stabilized by a second line of contacts parallel to the main diagonal, while the (anti-parallel) $\beta$-sheet  forms a series of contacts along the counter-diagonal.

\begin{figure*}[!h]
\begin{centering}
\includegraphics[width=17cm, height=8cm]{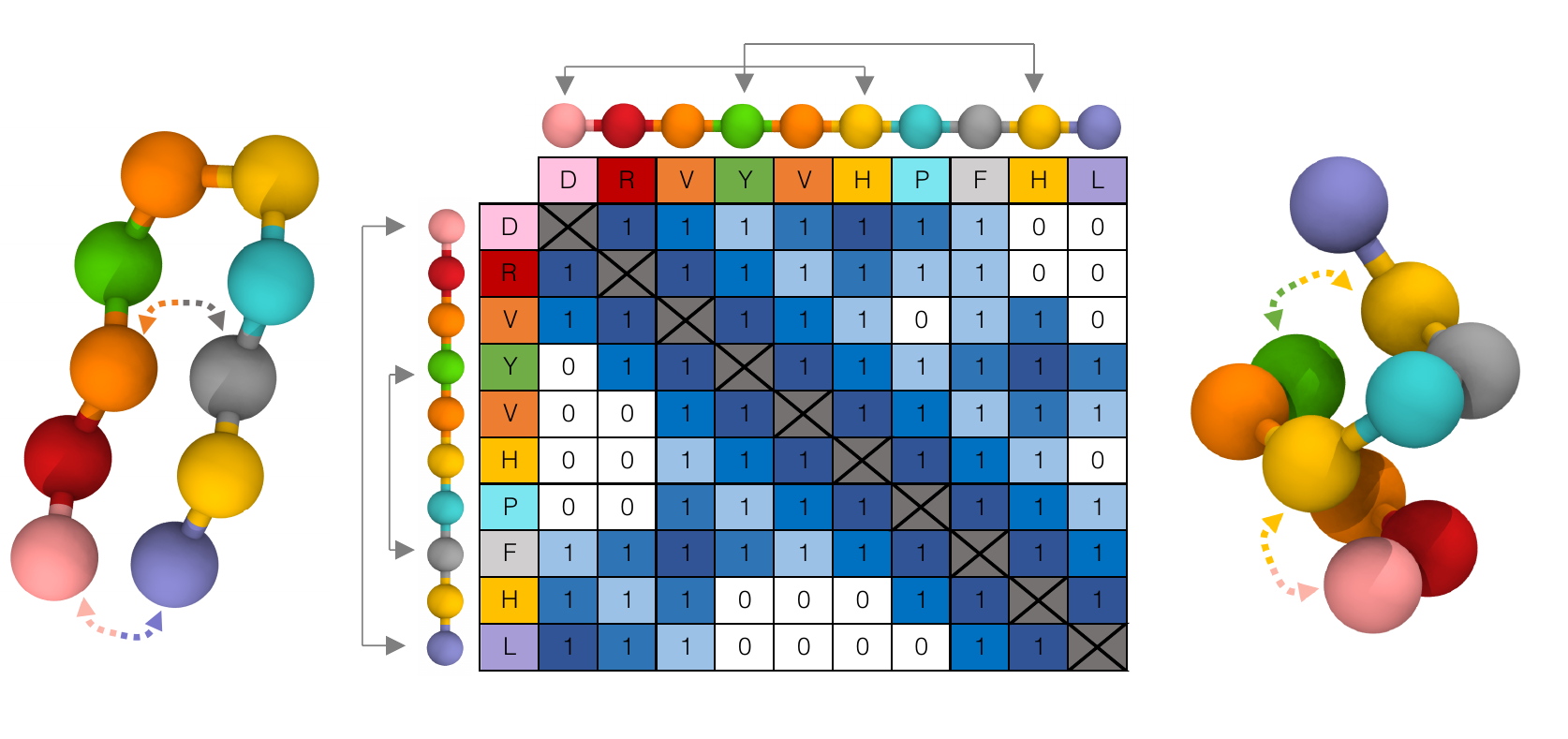}
\par\end{centering}
\caption{\textbf{Contact maps for the stabilization of secondary structure elements: $\beta$-sheet (left) and $\alpha$-helix (right).} 
The upper triangle of the contact map is designed to stabilize a $\alpha$-helix turn as showed in the structure on the right, which is obtained from a simulation based on the proposed model Hamiltonian.
On the other hand, the contacts in the lower triangle give rise to an anti-parallel $\beta$-sheet turn (structure to the left, also obtained from a simulation).
The signature of these two secondary structural elements (a series of parallel and anti-parallel contacts with respect to the diagonal  of the contact map) can be clearly identified. Color code: the blue color shades depict nearest, second-nearest and third-nearest neighbour interactions. 
The grey arrows above the sequences point to the stabilizing interactions. 
 \label{fig:alpha_beta}}
\end{figure*}

\section{Convergence of CVaR VQE for noisy simulation}

Simulation of the folding of the peptide Angiotensin using a realistic noise model for hardware. 

\begin{figure*}[!h]
\begin{centering}
\includegraphics[width=17cm, height=5cm]{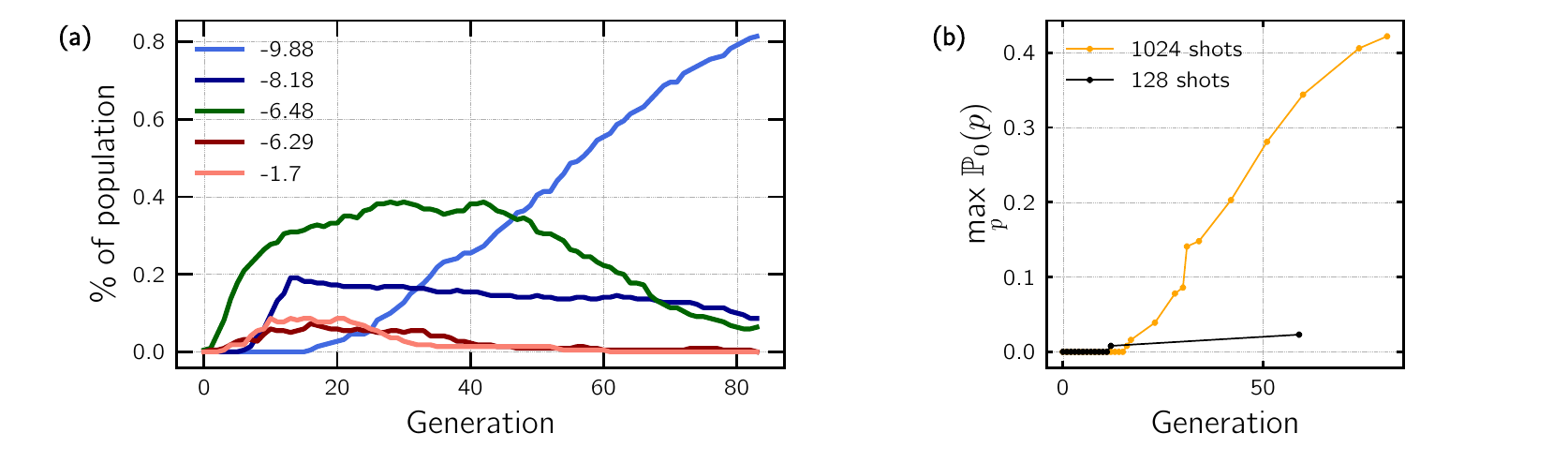}
\par\end{centering}
\caption{{(a) \textbf{Evolution of the ($\mathbb{P}_0(p) > 0$) throughout the minimization process for several low-energy conformations.} }
The corresponding conformations can be found in the main text.  
Results refer to the noisy simulations with 1024 shots (red bars in Figure 2 of the main article).
{(b) Same for $\max_p \mathbb{P}_0 (p)$ with $n_\mathrm{s}=1024$ and $n_\mathrm{s}=128$.} }
 \label{fig:alpha_beta}
\end{figure*}

\end{document}